\newcommand{\uu}[1]{\verb!#1!\endgroup}

\newcommand{\mb}[1]{\ifmmode#1\else\mbox{$#1$}\fi}

\newcommand{\fbar}      {\mb{\rm \bar f}}

\newcommand{\dbar}      {\mb{\rm \bar d}}
\newcommand{\ubar}      {\mb{\rm \bar u}}

\newcommand\al{\mb{\alpha}}

\newcommand\et{\mb{\eta}}

\newcommand\la{\mb{\lambda}}

\newcommand\Th{\mb{\Theta}}

\newcommand\Ph{\mb{\Phi}}

\newcommand\calM{\mb{{\cal M}}}

\newcommand{\slsh}{\rlap{$\;\!\!\not$}}                

\renewcommand{\Re}{{\cal R\!\rm e}}

\newcommand{\beq}{\begin{equation}}
\newcommand{\eeq}{\end{equation}}
\newcommand{\nn}{\nonumber}

\newcommand{\bea}{\begin{eqnarray}}
\newcommand{\eea}{\end{eqnarray}}

\newcommand{\fn}{\footnote}

\newcommand{\for}{\mb{{\rm for}}}
\newcommand{\emb}{\mb{{\rm emb}}}
\newcommand{\Ad}{\mb{\rm Ad}}

\newcommand{\tr}{\mb{\rm tr}}

\documentstyle[12pt]{article}
\topmargin = - 0.5 cm
\begin{document}
\bibliographystyle{unsrt}



\begin{flushright}
CERN-TH/95-97 \\
DAMTP-95-05
\end{flushright}
\begin{center}
\LARGE{EMBEDDED DEFECTS AND SYMMETRY BREAKING IN FLIPPED $SU(5)$\\}
\vspace*{0.5cm}
\large{Anne-Christine\ Davis
\fn{e-mail: A.C.Davis@amtp.cam.ac.uk}$^{,1,2}$
and Nathan\  F.\  Lepora
\fn{e-mail: N.F.Lepora@amtp.cam.ac.uk}$^{,2}$\\}
\vspace*{0.2cm}
{\small\em 1) Theory Division, CERN,\\
Geneva 23, CH-1211, Switzerland.\\}
\vspace*{0.2cm}
{\small\em 2) Department of Applied Mathematics and Theoretical
Physics,\\
University of Cambridge, Silver Street,\\
Cambridge, CB3 9EW, U.\ K.\\ }
\vspace*{0.2cm}

{February 1995}
\end{center}

\begin{abstract}
We explicitly show the analogy between the
symmetry breaking scheme for the GUT
flipped $SU(5)$ with that of
the Weinberg-Salam theory of electroweak interactions. This
allows us to construct the embedded
defect spectrum of the theory flipped $SU(5)$.
We find that the spectrum consists of
twelve gauge-equivalent unstable Lepto-quark strings, which are
analogous to W-strings in electroweak theory, and another string that
is gauge inequivalent to the Lepto-quark strings, which we call
the `V-string'. The V-string is analogous to the Z-string of
electroweak theory, correspondingly admitting a stable
semilocal limit. Using data on the running coupling constants we
indicate
that in the non-supersymmetric case V-strings can be stable for
part of the physically-viable parameter space. Cosmological
consequences are briefly discussed.
\end{abstract}
\thispagestyle{empty}
\newpage
\setcounter{page}{1}


\section{Introduction.}
\label{sec-intro}

Flipped $SU(5)$ \cite{Barr82}
is a very special Grand Unified Theory (GUT). It
stands out for its ease and simplicity of structure. Furthermore, many
of the problems associated with unification are simply not present for
this model. One of the practical problems of GUT's is their
testability; being physical theories they have to be testable and thus
falsifiable.
However, GUT's are at such an extreme energy scale that there
are few direct tests.
Hard predictions, such as proton decay, are few and far
between. Additionally, the generality of such predictions tends not to
discriminate between rival GUT's. Thus, one must start using arguments
such as naturalness and simplicity to focus attention onto your
favourite theory. These arguments are not physics, but they do serve
to motivate attention onto specific models where, perhaps with more
study, they may yield some interesting Cosmology.

For clarity of structure flipped $SU(5)$ stands proud. It is a very
simple theory and its close similarity in structure to the
Weinberg-Salam model of Electroweak interactions is a very notable
feature (we shall discuss this point later). It has a tiny Higgs
structure --- requiring only a 10-representation to facilitate
symmetry breaking
to the Standard Model. Furthermore, flipped $SU(5)$ is obtainable from
fundamental string theory~\cite{Lope93}.

Problems in physics often indicate new and exciting
structure. Generally, the number of such problems increase until a
crisis state is reached. The Standard model has a couple of problems
in the abstraction of its structure, but by no means is in a critical
state of affairs. The standard model's problems notably include the
monopole
 and coupling unification problem \cite{Lang93}. Flipped $SU(5)$
conveniently solves these problems; stable monopoles are not present
\cite{Dere84}
and the coupling
constants are not required to meet. Also, flipped $SU(5)$ has
a natural see-saw mechanism to guarantee light right-handed neutrinos
\cite{Elli90}.

Owing to the impossibility of directly testing GUT's in a terrestrial
experiment, one turns to the only situation where they were directly
relevant --- the Early Universe. Some GUT's yield strong, specific
Cosmological predictions. Cosmic strings being {\em the}
example \cite{Kibb76}.
However, there are no topological defects in flipped $SU(5)$,
rendering it seemingly untestable in this area. Fortunately this
happens
not to be the case --- it does yield non-topological defects, which
may be stable. Recent work on Z-strings in Electroweak theory
\cite{Vach92-1} and the
close formal similarity of Electroweak theory to flipped $SU(5)$
suggests that similar structures may be present for
flipped $SU(5)$. This work shows that there {\em are} embedded defects
in flipped $SU(5)$. Moreover, the counterpart of the Z-string in
flipped $SU(5)$ is very likely stable. Thus flipped $SU(5)$ should
yield a definite (and quite distinct) Cosmological signature. We shall
briefly indicate some Cosmological consequences in this paper. More
detailed calculations are presently underway.

In section~\ref{sec-flipped su5 : the model} we briefly review the
structure of flipped $SU(5)$, in such a way as to set the scene for a
detailed analysis of symmetry breaking. Symmetry breaking is
considered in section~\ref{sec-symmetry breaking in flipped su5}. Analysis
of the symmetry breaking facilitates a discussion of the embedded
defects structure, which is discussed in section~\ref{sec-embedded defects in
flipped su5}. We find unstable Lepto-quark strings and another
string that may be stable, this is called the V-string. Finally in
section~\ref{sec-discussion of results} we discuss the stability and
Cosmological consequences of this V-string and summarise our
conclusions.


\section{Flipped SU(5): The Model.}
\label{sec-flipped su5 : the model}

In this section we express previous work in flipped $SU(5)$ in a way
that will allow a detailed analysis of symmetry breaking and will thus
allow us to determine the embedded defect structure of the model. For
references see \cite{Barr82}, \cite{Dere84}, \cite{Elli90} and
\cite{Anto87}.

Flipped $SU(5)$ is a Yang-Mills gauge theory with a symmetry breaking
potential.
The grand unified gauge group is
$SU(5) \times \widetilde {U(1)}$. In the
 general case the
coupling constants of the simple groups, $g$ and $\tilde g$ respectively,
may be different --- as observed from the running couplings in
non-supersymmetric gauge theories. Generally, if the couplings do not
meet at Grand Unification then a non-simple gauge group is required;
such gauge groups are often inspired from Fundamental string theory.

In the following sections,
we adopt a notation where fields associated with the $\widetilde
{U(1)}$ part of the gauge group will be denoted with a tilde. This is
to make it plain which part of the group is being dealt with.

A modified Gell-Mann basis is used for $SU(5)$, as given in the
appendix. This basis is orthonormal with respect to the inner
product. The Lie algebra for the $\widetilde{U(1)}$ part is just a
phase and is proportional to the identity.

Hence, under its inner product, the Lie algebra
decomposes into the direct sum \mbox{$L(SU(5)) \oplus
L(\widetilde{U(1)}$}. Then
the gauge field is denoted by $A^\mu = A^\mu_a T_a + \widetilde {A^\mu}
\widetilde T$, where the sum in $a$ runs from 1 to 24.

The matter fields
transform as the representation $(n,q_n)$. Here $n$ specifies
the dimension of the representation of $SU(5)$ and $q_n$ is the
$\widetilde {U(1)}$-charge, which is defined by
\beq{
d_n(\widetilde T)(n,q_n)=i\sqrt{\frac{12}{5}} q_n(n,q_n),}
\eeq
with $d_n$ the derived $n$-dimensional representation of the Lie
algebra. Note the normalisation factor of $\sqrt{12/5}$
coming from the definition of
the $\widetilde{U(1)}$-generator.
Furthermore, an anomaly cancellation \cite{Dere84} yields $q_n$ as
$q_n=5-2m$, where $m$ is the
number of anti-symmetric indices labelling the components of the
$n$-representation.

The fermions are assigned to the following
representations of $SU(5)$:
the trivial 1-representation, the fundamental
5-representation and the antisymmetric 10-representation.
These transform under $SU(5)$ as, respectively:
\bea
D_1(g)M_1 &=& M_1,\ {\rm for}\ M_1 \in (1,5), \nn \\
D_5(g)M_5 &=& g M_5,\ {\rm for}\ M_5 \in (5,3), \\
D_{10}(g)M_{10} &=& g M_{10} g^{\rm T},
\ {\rm for}\ M_{10} \in (10,1). \nn
\eea
The conjugate representations have an $\widetilde {U(1)}$-charge of
opposite sign.

Fermions are assigned to these
representations flipped---${\rm u}\leftrightarrow {\rm d}$---
relative to the usual $SU(5)$ fermion assignments, so for the
left-handed fermions
\bea
{\rm f}^{10}_L&=& (10,1) =
\pmatrix{0&\dbar_L^1&-\dbar_L^2&\vdots&{\rm d}_L^1&{\rm u}_L^1\cr
                  -\dbar_L^1&0&\dbar_L^3 &\vdots&{\rm d}_L^2&{\rm u}_L^2\cr
                  \dbar_L^2&-\dbar_L^3&0&\vdots&{\rm d}_L^3&{\rm u}_L^3\cr
                  \ldots&\ldots&\ldots&\ldots&\ldots&\ldots\cr
                  -{\rm d}_L^1&-{\rm d}_L^2&-{\rm d}_L^3&\vdots&0&{\rm
   \overline \nu}_L\cr                  -
{\rm u}_L^1&-{\rm u}_L^2&-{\rm u}_L^3&\vdots&-{\rm \overline \nu}_L&0\cr},
\label{eq-b2}\\
\fbar^5_L &=& (\overline{5},-3) =
\pmatrix{\ubar_L^1\cr
                          \ubar_L^2\cr
                          \ubar_L^3\cr
                          \ldots\cr
                          {\rm \nu}_L\cr
                          {\rm \overline {e}}_L^+\cr},
\label{eq-b3} \\
{\rm f}^1_L &=& (1,5) = {\rm e}_L^-.
\label{eq-b4}
\eea
The right-handed fermions are assigned to the conjugate
representations. Fermions have been so assigned to have
the correct observed hypercharges (see
eq.~(\ref{eq-bhyp})). The flipped
representations are tied in to the non-trivial symmetry breaking,
which is further
tied in to the Higgs representation
chosen for symmetry breaking. Also
note the inevitable existence of a right-handed neutrino, which gives
rise to a see-saw mechanism \cite{Elli90}.

To achieve the desired symmetry breaking (see
section~\ref{sec-symmetry breaking in flipped su5}) a $(10,1)$
representation of the Higgs field is used.
An element of the Higgs field  is denoted
by the corresponding fermion assignment---$\overline \nu_
{\rm H}$ and so on.

The  Lagrangian for this model is written as
\beq{
{\cal L} = {\cal L}_f + {\cal L}_h + {\cal L}_g,}
\eeq
such that
\bea
{\cal L}_f &=& i\sum_n( \fbar^n_L \slsh D {\rm f}^n_L
                      + \fbar^n_R \slsh D {\rm f}^n_R), \nn \\
{\cal L}_h &=& {\tr}[(D_\mu \Phi)^\dagger D^\mu \Phi] - V(\Phi),
\label{eq-b5}\\
{\cal L}_g &=& {1 \over 4} {\tr}(F^a_{\mu \nu} F^{\mu \nu}_a) +
             {1 \over 4} {\tr}(\widetilde {F_{\mu \nu}}
              \widetilde {F^{\mu \nu}}), \nn
\eea
where the summation in ${\cal L}_f$ is over the different
 fermion representations. The Higgs potential, $V(\Phi)$, is
\beq{
V(\Ph) = \la_1({\tr}(\overline \Ph \Ph) - \et^2)^2 +
         \la_2 {\tr}(\overline \Ph \Ph \overline \Ph \Ph).
\label{eq-b6}}
\eeq
Also, the covariant derivative is written as
\beq{
D^\mu M_n = \partial ^\mu M_n + g d_n(A^\mu_a T_a) M_n +
            \widetilde g d_n(\widetilde {A^\mu} \widetilde T) q_n
            M_n.}
\eeq
Note that in the above Lagrangian there are no Yukawa coupling terms
for fermions to Higgs---this is because there are no gauge
invariant Yukawa coupling terms.


\section {Symmetry Breaking in Flipped $SU(5)$.}
\label{sec-symmetry breaking in flipped su5}

For $\et^2>0$ and $\lambda_1,(2\lambda_1 + \la_2)>0$, the Higgs
 potential (\ref{eq-b6}) has a set of
degenerate minima of the Higgs field
corresponding to the vacuum manifold. Furthermore,
for such ranges of the parameters the gauge group is transitive
over the vacuum manifold. Thus, the vacuum manifold (denoted by
$\calM_0$) can be written as
\beq{
{\cal M}_0 = \{ \Ph_c:V(\Ph_c) {\ {\rm is}\ {\rm a}\ {\rm minimum.}} \}
 \cong {SU(5) \times \widetilde {U(1)} \over {SU(3)
 \times SU(2) \times U(1)}}. }
\eeq

The process of symmetry breaking through a phase transition
is described by the familiar picture of
the Kibble mechanism \cite{Kibb76}.
The Higgs field tries to relax to the vacuum to minimise its potential
energy via taking a vacuum expectation value (VEV), $\Phi_c \in {\cal M}_0$.
 Due to the finite time over which the phase-transition takes
place $\Phi_c$ will not be uniform in space. Certain boundary
conditions arise naturally and the Higgs field will relax to a state
of minimum potential and kinetic energy. This leads to background
configurations, which may or may not be stable. We study the
configurations that occur in flipped $SU(5)$ in the next section.

 Since the gauge group is transitive
over the vacuum manifold, a gauge
rotation may be performed
on the VEV so that, without loss of generality, the Higgs VEV can be
in the ${\overline \nu}_H$ direction,
\beq{
\Ph_c = v \pmatrix{0&0&0&\vdots&0&0 \cr
               0&0&0&\vdots&0&0 \cr
               0&0&0&\vdots&0&0 \cr
               \ldots&\ldots&\ldots&\ldots&\ldots&\ldots\cr
               0&0&0&\vdots&0&1 \cr
               0&0&0&\vdots&-1&0 \cr}.
\label{eq-x1}}
\eeq
The size is
found by minimising the Higgs potential, giving
\beq{
v^2 = \frac{\la_1 \et^2}{2\la_1 + \la_2}.}
\eeq
Locally the VEV can always be
rotated to such a standard value. Although, globally such a situation
does not always exist. This is the origin of background configurations.

A short calculation using eq.~(\ref{eq-b5})
yields the mass terms for gauge bosons to be
\beq{
{\cal L}_{{\rm gauge}\ {\rm mass}}
             = {\tr}[(d_{10}(gA^\mu_aT_a+g'\widetilde {A^\mu}
             \widetilde T)\Ph_c)^\dagger
             (d_{10}(gA^\mu_aT_a+g'\widetilde {A^\mu}
             \widetilde T)\Ph_c)].
\label{eq-b7}}
\eeq
Now, $d_{10}(T_a)H_c = 0$ for $a=1..8$ and $a=22,23,24$. All other
generators (including $\widetilde T$) create mass terms for the gauge
fields. Thus,
one identifies colour-$SU(3)$ as being generated by $T_a :
a=1..8$ and isospin-$SU(2)$ as being generated by $T_a : a=22,23,24$.
Furthermore, analogously to the electroweak model,
there is a linear combination of generators that gives rise to another
massless gauge field, such that its generator lies perpendicular
to the vacuum. This generator is a linear combination of
$\widetilde T$ and $T_{15}$.

It should be noted that,
since $T_{15}$ is symmetric and $\Ph_c$ is antisymmetric,
then $d_{10}(T_{15})\Ph_c=2T_{15}\Ph_c$. Thus,
the minimal coupling of these generators to the Higgs vacuum is
\beq{
d_{10}(A^\mu)\Ph_c = 2 g A^\mu_{15} T_{15} \Ph_c +
                   {\tilde g} \widetilde {A^\mu} \widetilde T \Ph_c,
\label{eq-b8}}
\eeq
with $A^\mu = A^\mu_{15}T_{15} + {\widetilde A}^\mu \widetilde T$.

Analogy with the electroweak model yields the hypercharge generator, $T_Y$,
and a massive generator, $T_V$ (the analogue of the Z-boson
generator), by an orthogonal
rotation of $T_{15}$ and $\widetilde T$. To give the same minimal
coupling, with $A^\mu=A^\mu_a T_a + Y^\mu T_Y + V^\mu T_V$,
the corresponding rotation of the gauge fields is
\bea
{\left( \matrix {A^\mu_{15}\cr \widetilde {A^\mu} \cr} \right)} &=&
{\left( \matrix {\cos \Theta & -\sin \Theta \cr
                \sin \Theta & \cos \Theta \cr} \right)}
{\left( \matrix {Y^\mu \cr V^\mu \cr} \right)},
\label{eq-b9}\\
{\left( \matrix {g T_{15}\cr \tilde {g} \widetilde T \cr} \right)} &=&
{\left( \matrix {\cos \Theta & -\sin \Theta \cr
                \sin \Theta & \cos \Theta \cr} \right)}
{\left( \matrix {g_Y T_Y \cr g_V T_V \cr} \right)}. \nn
\eea
For hypercharge to be a gauge symmetry of the Standard Model,
$T_Y$ must be perpendicular to the vacuum,
$d_{10}(T_Y)\Ph_c=0$. This is so
only for
\beq{
\tan \Theta = \frac{\tilde g}{g}.}
\eeq
With this angle, squaring eq.~(\ref{eq-b9}) yields the couplings to
the hypercharge and V-bosons to be
\beq{
g_Y={\sqrt{{ g^2 \tilde {g}^2} \over { g^2 + \tilde {g}^2}}},
\hskip 1cm
g_V={\sqrt{{ g^4 + \tilde {g}^4} \over { g^2 + \tilde {g}^2}}}.
\label{eq-b10}}
\eeq
The gauge couplings to the Lepto-quark gauge bosons and
the gauge fields of $SU(2)$ and $SU(3)$ are $g$. One interprets
$\Theta$ as a generalised Weinberg angle appertaining to GUT's.

Thus the hypercharge generator and the V-generator are defined from
$T_{15}$ and $\widetilde{T}$ to be
\bea
T_Y &=& \frac{1}{\sqrt{2}}(\widetilde{T}+T_{15}) \\
T_V &=& \frac{\tilde{g}^2}{\sqrt{g^4 + \tilde{g}^4}}\widetilde{T}
- \frac{g^2}{\sqrt{g^4 + \tilde{g}^4}}T_{15} \nn
\eea

A convenient check on the above calculation is when $g=\tilde g$, then
necessarily $g_Y=g_V=g$.

To check the fermion assignments of
Section~\ref{sec-flipped su5 : the model}
(eqs. (\ref{eq-b2}), (\ref{eq-b3}) and (\ref{eq-b4})),
the hypercharges have to be verified to be correct. The
fermion hypercharges are given by the eigenvalues of the operator
$d_n(T_Y){\rm f}^n_L$. Using eq.~(\ref{eq-b10})
plus a little algebra this becomes
\beq{
id_n(T_Y){\rm f}^n_L={g\cos \Theta \over
{g_Y}}[d_n(T_{15})+\sqrt{\frac{12}{5}}(iq_n{\tilde {g}
\over g}\tan \Theta )]{\rm f}^n_L.}
\eeq
Using this the hypercharges of the fermions are
\bea
id_1(T_Y){\rm f}^1_L&=&-\sqrt{15} 2 \, {\rm f}^1_L, \nn \\
id_{\overline {5}}(T_Y) \fbar^5_L&=&
-\sqrt{15} {\rm diag}(-4/3,-4/3,-4/3;-1,-1) \, \fbar^5_L,
\label{eq-bhyp}\\
id_{10}(T_Y){{\rm f}^{10}_L}&=&
-\sqrt{15} {\rm diag}(2/3;1/3;0) \, {\rm f}^{10}_L; \nn
\eea
which verifies that the correct particle assignments have been made. There is a
normalisation factor $\sqrt{15}$
 which is included in the
definition of the coupling constant.

For completeness, we give details of the Higgs mechanism and the
masses of the bosonic sector of the theory in appendix B.


\section{Embedded Defects in Flipped $SU(5)$.}
\label{sec-embedded defects in flipped su5}

As we have seen from the last section, the flipped $SU(5)$ model is
very similar to the electroweak model in the pattern of symmetry
breaking. Both theories are of the form $SU(n) \times U(1)$ and break
to a group which has a $U(1)$ factor between the $SU(n)$ and the
$U(1)$ parts. In the electroweak model this structure gives a
non-trivial embedded defect structure; yielding W-strings
\cite{Vach92-2} and
Z-strings
\cite{Vach92-1}, which are stable in the semi-local limit
\cite{Vach91}. Thus, it seems
sensible to suppose these
structures also exist in
flipped $SU(5)$.
Furthermore, it is known that in the electroweak model Z-strings
are unstable for physical Weinberg angle
\cite{Jame92}. However, for GUT's the
parameters are different and the embedded defect might well be
stable. This situation is particularly favourable in flipped $SU(5)$
owing to the coupling constants not being required to meet at
unification. This situation is analysed in the next section.

We shall show that for flipped $SU(5)$ there are two classes of gauge
inequivalent embedded Nielsen-Olesen solutions. One class contains one
element (the analogue of the Z-string); the other class contains
twelve elements (the analogues of the W-strings). We refer to
the one-dimensional class of embedded vortices as `V-strings' ---
because these are generated by the generator that gives the V-boson.

The stability of such embedded defects is a very important issue and
we shall show the analogy between the Z-string of electroweak theory
and the V-string may be carried further; namely that the V-string is
stable in an appropriate semi-local limit.

\subsection{Existence of Embedded Defects in Flipped $SU(5)$.}
\label{existence}

We follow the approach of \cite{Vach94} for
describing the existence of embedded defect solutions

To describe the embedded solutions we need a reference point on the
vacuum manifold in order to generate the solution and to describe
different, but possibly gauge equivalent, solutions. We take this
point to be, without loss of generality, $\Phi_c$ as given by
eq.~(\ref{eq-x1}). For
convenience we also need a basis for $L(SU(5) \times {\widetilde {U(1)}})$
orthonormal with respect to the natural Inner Product;
we take this basis to be as
given in the appendix.

To define the embedded solutions one considers one-parameter
subgroups $G^{\emb} \subseteq SU(5) \times {\widetilde {U(1)}}$, such
that its corresponding little group $H^{\emb} = G^{\emb} \cap \{
SU(3)_C \times SU(2)_I \times U(1)_Y \}$ is trivial. These
one-parameter groups are generated by broken generators, i.e. those
generators which satisfy
\beq{
d_{10}(T_a) \Phi_c \not= 0.}
\eeq
We denote this vector space of broken generators by $M$ and it is
spanned by the Lepto-Quark generators and by the V-boson generator.
In the basis previously mentioned
\beq{
M = {\rm span} \{T_V, T_a : a=9..14, 16..21 \}.}
\eeq
Then for each $T_s \in M$ one has a one-parameter subgroup of
$SU(5) \times {\widetilde {U(1)}}$, given by
\beq{
G^{\emb}[T_s] = \{g(\theta)={\rm exp}
\left( {\theta T_s \over {c(T_s)}} \right)
: \theta \in [0,2\pi) \},}
\eeq
where $c(T_s)$ is a normalisation constant such that
$g(2\pi)={\rm id}_G$ and
there does not exist a
$\theta_0 \in (0, 2\pi)$ with $g(\theta_0) = {\rm id}_G$
(we are here denoting the identity of the gauge group by ${\rm id}_G$).

We then define an embedded subspace of the representation-space,
${\cal V}^{\emb}[T_s] \subseteq {\cal V}$ by
\beq{
{\cal V}^{\emb}[T_s] = \{ \alpha {\rm exp} \left({\theta d_{10}(T_s)
\over {c(T_s)}} \right)
\Phi_c : \theta \in [0,2\pi), \alpha \in \Re \}.}
\eeq

It is then clear that, under this construction, ${\cal V}^{\emb}
[T_s]$ is invariant under the action of $G^{\emb}[T_s]$. Hence,
we have constructed a unique embedded sub-theory from the element $T_s
\in M$, that is defined by $(G^{\emb}[T_s],{\cal
V}^{\emb}[T_s])$. This embedded sub-theory has the property that its
little group  $H^{\emb} = G^{\emb} \cap \{
SU(3)_C \times SU(2)_I \times U(1)_Y \}$ is trivial.

The vacuum manifold for the embedded sub-theory is defined to be
\beq{
{\cal M}^{\emb}_0[T_s] = {\cal M}_0 \cap {\cal V}^{\emb}[T_s]}
\eeq
and one sees that all such embedded vacuum manifolds have non-trivial
first homotopy groups, since
\beq{
{\cal M}^{\emb}_0[T_s] \cong {G^{\emb}[T_s] \over {H^{\emb}[T_s]}}
\cong S^{(1)}.}
\eeq

If one considers just the embedded sub-theory in isolation,
then it
exhibits Nielsen-Olesen solutions due to its non-trivial
topological nature. These
embedded solutions are given by (in the temporal gauge):
\bea
\Phi(r, \theta) &=& f_{\rm NO}(r) D_{10}({\rm exp}({T_s \theta \over
{c(T_s)}}))\Phi_c, \nn \\
\underline A &=& {g_{\rm NO}(r) \over {g_s r}} \left( {T_s \over {c(T_s)}}
\right) \underline{\widehat \theta}, \\
A^0 &=& 0. \nn
\eea
Here $f_{\rm NO}(r)$ and $g_{\rm NO}(r)$ are the profile functions for a
Nielsen-Olesen solution and satisfy
\bea
f''_{\rm NO} + {f'_{\rm
NO} \over r} - {(1 + g_{\rm NO})^2 \over r^2}f_{\rm NO}
&=& -f_{\rm NO}(m_1^2 + 2(\lambda_1 + \lambda_2/2)f^2_{\rm NO}), \\
({-1 \over {c(T_s)^2}})(g''_{\rm NO} - {g'_{\rm NO} \over r}) &=&
-4g_S^2 v^2 (g_{\rm NO} +1)f_{\rm NO}^2, \nn
\eea
where this equation has been derived from the equations of motion.

When this embedded solution is taken back to the full theory
it still remains a solution provided that
\beq{
{(} \Phi (r,\theta), {\cal V}^\perp {)} = 0,}
\eeq
where ${\cal V}^\perp = \{ \Phi \in {\cal V} : (\Phi, {\cal V}^{\emb})=0
\}$ and
\beq{
(d_{10}(A^\mu),d_{10}(T_s^\perp))=0,}
\eeq
with $T_s^\perp = \{T \in L(SU(5) \times {\widetilde {U(1)}}) : (T,
T_s) =0\}$.

It can be shown \cite{Vach94}
that these conditions are generally satisfied if
\beq{
d_{10}(T_s)\Phi \in {\cal V}^\perp,{\rm \ for\ }\Phi \in {\cal V} {\rm
\ and \ }T \in T_s^\perp},
\eeq
which we can verify is satisfied for flipped $SU(5)$.

Before writing down the embedded solution we should firstly derive
the condition for two different embedded sub-theories (defined by, say,
$T_s$ and $T_s'$) to be gauge equivalent. We shall deal with gauge
transformations of the type
\bea
\Phi &\mapsto& D_{10}(g)\Phi, \nn \\
A^\mu &\mapsto& {\Ad}(g) A^\mu, \\
{\rm such\  that}\ g &\in& SU(3)_C \times SU(2)_I \times U(1)_Y.\nn
\eea

A short calculation shows that this is equivalent to a rotation of the
generator, $T_s$, which defines the sub-theory, of the form
\beq{
T_s \mapsto {\rm Ad}(g)T_s.}
\label{eq-bAd}
\eeq
In other words
\beq{
T_s \mapsto e^T T_s e^{-T},\ {\rm with \ } T \in
L(SU(3)_C) \oplus L(SU(2)_I) \oplus L(U(1)_Y).}
\eeq
Therefore, two embedded solutions (with string generators $T_s$ and
$T_s'$, say) are gauge equivalent if there
exists a $g \in SU(3)_C \times SU(2)_I \times U(1)_Y$ such that $T_s'
= D_{10}(g) T_s$.

For flipped $SU(5)$ it is simple, but tedious, to verify that the
generators of the sub-theory, $T \in M$, split into two gauge
inequivalent classes under eq.~(\ref{eq-bAd}). Then
\bea
M &=& M_1 \oplus M_2, \nn \\
{\rm with\ } M_1 &=& {\rm span} \{T_V\}, \\
{\rm and\ } M_2 &=& {\rm span} \{T_a : a=9..14,16..21\}, \nn
\eea
and for $T,T' \in M_2$ there exists a
$g \in SU(3)_C \times SU(2)_I \times
U(1)_Y$ such that $T = D_{10}(g)T'$.

Hence, for flipped $SU(5)$ there are two gauge inequivalent classes of
solution. Firstly, the V-string solution
\bea
\Phi(r, \theta) &=& f_{\rm NO}(r) D_{10}({\rm exp} \left( {T_V \theta \over
{c(T_V)}}\right) )\Phi_c = f_{\rm NO}(r) e^{i\theta} \Phi_c, \nn \\
\underline A &=& {g_{\rm NO}(r) \over {g_V r}}
\left( {T_V \over {c(T_V)}} \right)
\underline{\widehat \theta}, \\
A^0 &=& 0 \nn
\eea
with $c(T_V)$ a calculable constant dependent upon gauge couplings.
The other class of solutions
are the Lepto-Quark strings and are generated by $T_{\rm LQ} \in M_2$,
having the solution
\bea
\Phi(r, \theta) &=& f_{\rm NO}(r) D_{10}({\rm exp}
\left( {T_{\rm LQ} \theta \over
{c(T_{\rm LQ})}}\right) )\Phi_c, \nn \\
\underline A &=& {g_{\rm NO}(r) \over {g_{\rm LQ} r}}
\left( {T_{\rm LQ} \over {c(T_{\rm LQ})}} \right)
\underline{\widehat \theta}, \\
A^0 &=& 0. \nn
\eea
and as shown previously all Lepto-Quark strings are gauge equivalent.

\subsection{The Semi-Local Limit of the V-string.}
\label{semi-local}

That there was a resurgence of interest in the Z-string of electroweak
theory was due to the fact that it is stable in the semi-local limit
of the Weinberg-Salam model
\cite{Vach91}, but not for physical Weinberg angle \cite{Jame92}.
Taking the analogy between flipped $SU(5)$ and electroweak
theory, one would expect that the V-string should be stable in the
semi-local limit. Furthermore, the domain of stability {\em might} overlap
the point of physical reality in a GUT scale scenario --- this
is discussed in the next section.

Work by Preskill in \cite{Pres92} has shown how to identify stable
semi-local defects when one considers a general
symmetry breaking $G \rightarrow H$.
Consider a subgroup $G_{\rm gauge} \subseteq G$ and $H_{\rm gauge} =
G_{\rm gauge} \cap H$. If one considers $G$ to be a global symmetry
(zero gauge-coupling) and then gauges just $G_{\rm gauge}$, the
existence of semi-local defects is indicated by the non-trivial homotopy
classes of $G_{\rm gauge}/H_{\rm gauge}$. Semi-local vortices,
which  have a
corresponding non-trivial first homotopy class, are stable.

Hence for flipped $SU(5)$ the appropriate semi-local limit of the
model is when $\Th_{\rm GUT} \rightarrow \frac{\pi}{2}$, so that just
the $\widetilde{U(1)}$ symmetry is gauged. Since $\widetilde{U(1)}
\cap H = {\rm id}_G$, the topology of the situation gives stable
semi-local defects generated by $\widetilde{T}$. It is clear that
these semi-local defects correspond (with a gauge transformation) to
the V-strings previously discussed.


\section{Discussion of Results.}
\label{sec-discussion of results}

Owing to extremely accurate measurements of the Z mass, the electroweak
Weinberg angle, electric charge and strong charge, the low energy
coupling constants for $SU(3)_c$, $SU(2)_I$ and $U(1)_Y$ are known to
an unprecedented degree of accuracy. Using renormalisation group
techniques the high energy values of these couplings can be
calculated. GUT's, in general, have some constraints on the form of
couplings and also have a lower bound upon the unification energy
scale (from proton decay rates). We are thus in a position to apply some
physics to the existence of stable V-strings. The results on running
couplings are taken from \cite{Lang93}.

\subsection{Coupling Constant Unification.}
\label{subsec-coupling constant unification}

For GUT's that have simple gauge groups, coupling constants are
required to meet at unification. However, GUT's with a non-simple
gauge group are only constrained such that the strong and weak
couplings must meet. Thus from calculations of running couplings one
may determine the unification scale. Furthermore, from
the value of the hypercharge coupling constant at unification one may
calculate the value of ${\tan}\Th_{\rm GUT}$, which is the quantity
relevant for stability of V-strings.

Denoting the hypercharge coupling by $\al_1$, the weak (isospin) coupling
by $\al_2$ and the strong coupling by $\al_3$, renormalisation group
calculations yield the graph for running couplings in \cite{Lang93}.

The unification scale is determined from the energy when $\al_2 =
\al_3$, thus for flipped $SU(5)$ it is identified to be
\beq{
\mu_{\rm GUT} = 10^{16}\ {\rm to }\ 10^{17} {\rm GeV}.}
\eeq
and at this scale the coupling constants are
\beq{
\al^{-1}_{2 {\rm GUT}} = \al^{-1}_{3 {\rm GUT}} = 45\ {\rm to}\ 49,
\ \ \ \ \al^{-1}_{1 {\rm GUT}} = 36 \ {\rm to}\ 38.}
\eeq

To calculate the value of ${\tan}\Th_{\rm GUT}$ one needs the value of
the $SU(5)$ coupling constant, $g$ and the $\widetilde{U(1)}$ coupling
constant $\tilde g$. These can be easily calculated from $\al_{i {\rm
GUT}}$, then a simple calculation yields
\beq{
{\tan}\Th_{\rm GUT} = \frac{\tilde{g}}{g} = 1.2 \ {\rm to}\ 1.4.}
\eeq

The main uncertainty is in the value of the strong coupling
constant. However, the above quantities are enough for some simple
calculations.

The non-supersymmetric version of flipped $SU(5)$,
where the coupling constants do not meet, necessarily
precludes the existence of a further unification to one coupling
constant (i.e. a simple group) at a higher energy-scale. This is
because $SU(5)$ is asymptotically free and the $\widetilde{U(1)}$ is
not. Thus after unification the coupling constant diverge, precluding
further unification into a simple group. This is a generic feature of
any unification scheme of a similar form to
non-supersymmetric flipped $SU(5)$.

\subsection{Stability of V-strings.}
\label{subsec-stability of v strings}

We have shown that the V-string has a stable semi-local limit. The
stability of the corresponding semi-local vortex is dependant upon the
value $\la = (\la_1 + \la_2/2)$, which parameterises the strength of the
potential. For $\la \in (0,1)$ the semi-local vortex is stable and for
$\la \geq 1$ the vortex is unstable.

There is another line of interest in the $(\la, \Th_{\rm
GUT})$-stability plane, which is for $(\la, \Th_{\rm GUT}= \pi/4)$, where
the embedded vortex is unstable \cite{Vach94}.

The region of stability is separated from the region of instability by
a critical curve going from $(\la = 1, \Th_{\rm GUT} = \pi/2)$ to
$(\la =0, \Th_{\rm GUT}= \Th_{\rm crit})$, with $\Th_{\rm crit}
\in [\pi/4, \pi/2)$. To see an example of such a stability region, one
should refer to \cite{Jame92}, which calculates
 the form of the stability region for electroweak
Z-strings of the Weinberg-Salam model.
Note that $\Th_{\rm crit}$ is not obtained from the numerical
techniques used to evaluate these graphs. It is an open question what
the value of $\Th_{\rm crit}$ is. Another stability analysis, for the
case of the two-doublet Electroweak model, has been performed in
\cite{Earn93}.

 From comparing the results in \cite{Earn93} and \cite{Jame92},
it is clear that as the number of Higgs degrees of
freedom increase, then the region of stability gets larger. The
physical reason for this is related to observing that stability
depends upon whether the potential decreases more than the kinetic
terms upon small perturbations in the Higgs field. This is why
embedded defects are unstable for large $\la$. Having more Higgs
spreads the perturbation over more degrees of freedom --- having no
effects on changes in the potential, but causing the kinetic terms to
increase more. Thus more Higgs degrees of freedom increases the region
of stability.

Hence, for flipped $SU(5)$ it seems likely (or at least an open
question) that it has stable embedded defects. Firstly, because
flipped $SU(5)$, being a GUT, has many Higgs degrees of
freedom. Secondly, because $\Th_{\rm GUT}$ is large (certainly larger
than $\pi/4$).

\subsection{The Cosmology of Flipped $SU(5)$.}
\label{subsec-cosmology of flpped su5}

First of all, we point out that the unification energy scale is
compatible with present proton-decay rates. Thus flipped $SU(5)$
remains a viable option.

It is a well known feature of flipped $SU(5)$ that, provided it is not
embedded in a simple gauge group, there are no topological
monopoles. There are, however, embedded monopole solutions --- but these
are unstable. Thus, the monopole problem is circumvented in
flipped $SU(5)$.

The presence of stable embedded defects gives many cosmological
implications. We firstly describe the effect of cooling upon these
defects before considering general cosmological implications.

The first point to note for the cosmology is that the probability of
formation (per unit volume) is less for an embedded defect than for an
Abelian topologically stable defect. For an Abelian defect the
Kibble-mechanism \cite{Kibb76} says that we only need to consider
random phases in the $U(1)$ factor of the gauge group. However, for an
embedded defect the Kibble mechanism would predict that some
combination of a Lepto-quark string and a V-string would be initially
formed. This configuration is not a stationary point of the Lagrangian
and so some currents must be present to compensate for this. This
combination would form probabilistically
either a Lepto-quark string or a V-string. Thus the probability of
formation of a V-string must be less than that for an Abelian
string. The probability of formation
is also decreased by the region of stability \cite{Davi94}

As the universe cools the coupling constants run and ${\tan}\Th_{\rm
GUT}$ decreases. Hence, it appears that a stable embedded defect would
destabilise at a lower energy scale. This does not appear to be the
case. Observe that in the centre of the defect symmetry is restored and
hence it is this $\tan{\Th_{GUT}}$ that is relevant to stability and
not the low temperature value.
 This should stop a stable
embedded defect from decaying
due to the gauge couplings running as the temperature falls.

However, embedded defects are not topologically stable; they can only
be dynamically stable. This means there is no topological charge
to guarantee the lasting stability of such a defect and in general
such a defect will decay by the nucleation of an embedded monopole and
anti-monopole pair along the string \cite{Pres93}. A string ending
in a monopole and anti-monopole will then shrink due to tension along
it. It is an open question whether a short length of
dynamically stable string ending at such monopole pairs would be
stable or not. If such a configuration were completely stable then it
might over-close the Universe --- resulting in a cosmological
disaster!

The rate of creation of monopole and anti-monopole pairs should be
fairly high on a length of V-string. As the parameters of an embedded
defect get closer to the region of dynamical instability then the
probability of nucleation of monopole and anti-monopole pairs gets
higher. Thus, it is unlikely that a length of V-string would survive
until today. However, the rate of nucleation may be
low enough to ensure that stable configurations consisting
of short lengths of V-string ending in monopoles may be in quite small
abundance. This abundance could be small enough to circumvent the
over-closure problem.

Gravitational interactions of a stable embedded defect are the same as
topologically stable defects. Thus, if V-strings were to survive into
the matter-dominated epoch of the Universe then they would seed
structure formation. Also, a V-string would produce gravitational
lensing effects upon light travelling past it. Due to the (reasonably)
short lifetime of a V-string it is unlikely that it would survive long
enough to produce these effects. However, they may survive long enough
to pass through the surface of last scattering --- leaving a signature
upon the Cosmic microwave background. This last effect is probably the
only way of observing such strings directly.

If V-strings were able to live to the Electroweak phase transition
they should give Electroweak
baryogenesis. It has been recently shown that topological strings, in
passing through the Electroweak phase-transition, give baryogenesis
with the bias given by $CP$ \cite{Bran94}. A V-string has the same
interaction with matter as an Abelian topological string and thus they
would produce a similar effect.

It is clear from the above that the V-string, if it were stable, could
be Cosmologically very interesting. The crucial question is, however,
how long do they live? Too short a lifetime and their Cosmological
consequences could be minimal (or fatal if embedded monopole and
anti-monopole pairs joined by short V-strings were stable). A long
lifetime and then we have a realistic GUT producing strings with lots
of nice Cosmological effects.

We should point out that if stable V-stings were to be observed (or
indeed stable GUT-scale embedded defects of any form) then this would
contradict supersymmetry. Supersymmetry causes the
coupling constants to meet at unification --- meaning that any
GUT-scale embedded
defect will be unstable.

It is worth noting that if a Cosmological consequence turns out to be
particularly dire (such that it rules out our Universe)  then the
particular model that gives rise to this consequence is ruled out (or
revamped with some extra/different parameters).
Thus the consequences that we
have sketched could be fatal for flipped $SU(5)$.
So with all the glories and nice features of
flipped $SU(5)$, it is possible that embedded defects could
rule it out.

\bigskip
\bigskip


{\noindent{\LARGE{\bf Acknowledgements.}}}

\nopagebreak

\bigskip

\nopagebreak

This work is supported in part by PPARC. One of us, NFL, acknowledges
EPSRC for a research studentship. We thank N. Manton, M. Trodden and
T. Vachaspati for discussions.

\bigskip
\bigskip


{\noindent{\LARGE{\bf Appendix A.}}}

\nopagebreak

\bigskip

\nopagebreak

The modified Gell-Mann basis for $L(SU(5))$ is \cite{Bail}
\[
T_a = \frac{i}{\sqrt 2} \mu_a .
\]
The Inner Product on this vector space is given by
$(T_a,T_b)=tr(T_a^\dagger T_b)$. In order for $T_a$ to be orthonormal
with respect to this basis, $\mu_a$ with $a=1..24$ is defined to be:
\scriptsize
\bea
\mu_1 = \left( \begin{array}{ccccc} 0&1&0&0&0 \\
               1&0&0&0&0 \\
               0&0&0&0&0 \\
               0&0&0&0&0 \\
               0&0&0&0&0 \end{array} \right),\ \ \ \ \ \ \ \ \ \
\mu_2 &=& \left( \begin{array}{ccccc} 0&-i&0&0&0 \\
               i&0&0&0&0 \\
               0&0&0&0&0 \\
               0&0&0&0&0 \\
               0&0&0&0&0 \end{array} \right),\ \ \ \ \ \ \ \ \
\mu_3 = \left( \begin{array}{ccccc} 1&0&0&0&0 \\
               0&-1&0&0&0 \\
               0&0&0&0&0 \\
               0&0&0&0&0 \\
               0&0&0&0&0 \end{array} \right),\nn \\
\mu_4= \left( \begin{array}{ccccc} 0&0&1&0&0 \\
               0&0&0&0&0 \\
               1&0&0&0&0 \\
               0&0&0&0&0 \\
               0&0&0&0&0 \end{array} \right), \ \ \ \ \ \ \ \ \
\mu_5 &=& \pmatrix{0&0&-i&0&0 \cr
               0&0&0&0&0 \cr
               i&0&0&0&0 \cr
               0&0&0&0&0 \cr
               0&0&0&0&0 \cr},\ \ \ \ \ \ \ \ \
\mu_6=\pmatrix{0&0&0&0&0 \cr
               0&0&1&0&0 \cr
               0&1&0&0&0 \cr
               0&0&0&0&0 \cr
               0&0&0&0&0 \cr}, \nn \\
\mu_7 =\pmatrix{0&0&0&0&0 \cr
               0&0&-i&0&0 \cr
               0&i&0&0&0 \cr
               0&0&0&0&0 \cr
               0&0&0&0&0 \cr},\ \ \ \ \ \ \ \ \
\mu_8&=&{1 \over \sqrt 3}\pmatrix{1&0&0&0&0 \cr
               0&1&0&0&0 \cr
               0&0&-2&0&0 \cr
               0&0&0&0&0 \cr
               0&0&0&0&0 \cr},\ \ \ \ \ \ \ \ \
\mu_9 =\pmatrix{0&0&0&1&0 \cr
               0&0&0&0&0 \cr
               0&0&0&0&0 \cr
               1&0&0&0&0 \cr
               0&0&0&0&0 \cr},\nn \\
\mu_{10}=\pmatrix{0&0&0&-i&0 \cr
               0&0&0&0&0 \cr
               0&0&0&0&0 \cr
               i&0&0&0&0 \cr
               0&0&0&0&0 \cr},\ \ \ \ \ \ \ \ \
\mu_{11} &=&\pmatrix{0&0&0&0&0 \cr
               0&0&0&1&0 \cr
               0&0&0&0&0 \cr
               0&1&0&0&0 \cr
               0&0&0&0&0 \cr},\ \ \ \ \ \ \ \ \
\mu_{12}=\pmatrix{0&0&0&0&0 \cr
               0&0&0&-i&0 \cr
               0&0&0&0&0 \cr
               0&i&0&0&0 \cr
               0&0&0&0&0 \cr}, \nn \\
\mu_{13}= \pmatrix{0&0&0&0&0 \cr
               0&0&0&0&0 \cr
               0&0&0&1&0 \cr
               0&0&1&0&0 \cr
               0&0&0&0&0 \cr},\ \ \ \ \ \ \ \ \ \ \
\mu_{14}&=&\pmatrix{0&0&0&0&0 \cr
               0&0&0&0&0 \cr
               0&0&0&-i&0 \cr
               0&0&i&0&0 \cr
               0&0&0&0&0 \cr}, \ \ \ \ \
\mu_{15} ={\sqrt {3 \over 5}}
               \pmatrix{2/3&0&0&0&0 \cr
               0&2/3&0&0&0 \cr
               0&0&2/3&0&0 \cr
               0&0&0&-1&0 \cr
               0&0&0&0&-1 \cr},\nn \\
\mu_{16}=\pmatrix{0&0&0&0&1 \cr
               0&0&0&0&0 \cr
               0&0&0&0&0 \cr
               0&0&0&0&0 \cr
               -1&0&0&0&0 \cr}, \ \ \ \ \ \ \ \ \
\mu_{17} &=&\pmatrix{0&0&0&0&-i \cr
               0&0&0&0&0 \cr
               0&0&0&0&0 \cr
               0&0&0&0&0 \cr
               i&0&0&0&0 \cr},\ \ \ \ \ \ \ \ \
\mu_{18}=\pmatrix{0&0&0&0&0 \cr
               0&0&0&0&1 \cr
               0&0&0&0&0 \cr
               0&0&0&0&0 \cr
               0&1&0&0&0 \cr}, \nn \\
\mu_{19}= \pmatrix{0&0&0&0&0 \cr
               0&0&0&0&-i \cr
               0&0&0&0&0 \cr
               0&0&0&0&0 \cr
               0&i&0&0&0 \cr},\ \ \ \ \ \ \ \ \
\mu_{20}&=&\pmatrix{0&0&0&0&0 \cr
               0&0&0&0&0 \cr
               0&0&0&0&1 \cr
               0&0&0&0&0 \cr
               0&0&1&0&0 \cr},\ \ \ \ \ \ \ \ \
\mu_{21} =\pmatrix{0&0&0&0&0 \cr
               0&0&0&0&0 \cr
               0&0&0&0&-i \cr
               0&0&0&0&0 \cr
               0&0&i&0&0 \cr},\nn \\
\mu_{22}=\pmatrix{0&0&0&0&0 \cr
               0&0&0&0&0 \cr
               0&0&0&0&0 \cr
               0&0&0&0&1 \cr
               0&0&0&1&0 \cr}, \ \ \ \ \ \ \ \ \
\mu_{23} &=&\pmatrix{0&0&0&0&0 \cr
               0&0&0&0&0 \cr
               0&0&0&0&0 \cr
               0&0&0&0&-i \cr
               0&0&0&i&0 \cr},\ \ \ \ \ \ \ \ \
\mu_{24}=\pmatrix{0&0&0&0&0 \cr
               0&0&0&0&0 \cr
               0&0&0&0&0 \cr
               0&0&0&1&0 \cr
               0&0&0&0&-1 \cr}. \nn
\eea
\normalsize

\bigskip
\bigskip


{\noindent{\LARGE{\bf Appendix B.}}}

\nopagebreak

\bigskip

\nopagebreak

The Higgs fields gain mass via  their coupling
to the vacuum manifold. They do this by `eating' components of the
Higgs field---this can be seen by transforming to the unitary gauge
where it is transparent that
longitudinal polarisation component of the massive gauge fields are
from the Higgs degrees of freedom. The eaten components of the Higgs
field correspond to fields that are transverse to the vacuum
manifold. The rest of the Higgs field---which correspond to massive
Goldstone bosons---is from the radial Higgs fields $\Phi_R$, such that
\bea
\Phi_R &=& \{ \Phi_R=(\Phi - \Phi_c) : {\tr}(\Phi^\dagger_R
d_{10}(T)\Phi_c)=0, \\
& & \ \ \ \ \ \ \ \ \ \ \ \ \ \ \ \ \
T=T_a,T_V,T_V,{\for}\ a=1..23\}.\nn
\eea
This corresponds to a choice of generalised polar
coordinates in specifying the Higgs field. A short calculation yields
\beq{
\Phi_R=\pmatrix{0&\overline{d}_H^1&-\overline{d}_H^2&\vdots&0&0\cr
                  -\overline{d}_H^1&0&\overline{d}_H^3&\vdots&0&0\cr
                  \overline{d}_H^2&-\overline{d}_H^3&0&\vdots&0&0\cr
                  \ldots&\ldots&\ldots&\ldots&\ldots&\ldots\cr
                  0&0&0&\vdots&0&\sigma_H\cr
                  0&0&0&\vdots&\sigma_H&0\cr},}
\eeq
with $\overline{\nu}_H=\sigma_H e^{i\theta}$ and $\sigma_h \in
{\Re}$. The other components of the Higgs field
are represented by a gauge rotation of this,
yielding
\beq{
\Phi=\Phi_c + d_{10}(g)\Phi_R,}
\eeq
where $g \in SU(5) \times \widetilde {U(1)}$.

To obtain the mass terms for gauge bosons and Higgs one
substitutes $\Phi$ in ${\cal L}_h$ (\ref{eq-b5}) . The unitary gauge
is implicitly used to transform away transverse components of the Higgs
field, obtaining
\bea
{\cal L}_h &=& {\cal L}_{{\rm gauge}\ {\rm mass}}
           + {\cal L}_{{\rm Higgs}\ {\rm mass}} +{\cal L}_I,\nn \\
{\cal L}_{{\rm gauge}\ {\rm mass}} &=& g^2v^2 \sum^3_{i=1}
    ({\overline X}^{\mu}_i X_{i\mu}+{\overline Y}^{\mu}_i Y_{i\mu})
    +{32 \over {35}}g^2_Yv^2 {\overline V}^{\mu} V_{\mu},\\
{\cal L}_{{\rm Higgs}\ {\rm mass}} &=& -(m^2_1+8v^2)({\lambda_1 \over 2}
    +\lambda_2)(\sum^3_{i=1} {\overline u}_{iH} u_{iH} + {\overline
     \sigma}_H \sigma_H),\nn
\eea
with ${\cal L}_I$ being the interaction between gauge and Higgs
fields. The Lepto-quark bosons are $X^\mu,Y^\mu$, as described below.
Hence the masses of the respective particles are
\beq{
m^2_X = m^2_Y = g^2v^2,\
m_V^2 = {32 \over {35}}g^2_Yv^2,\
m_H^2 = -(m^2_1+8v^2)({\lambda_1 \over 2}
    +\lambda_2).}
\eeq
Note that since $\lambda_1,\lambda_2 \sim 1$ and $-m^2_1 \sim
v^2$, the coefficient of the Higgs mass term is positive.

For completeness we shall describe which Higgs fields are eaten by
which gauge bosons. To make this explicit, choose a basis where the
Lepto-Quark gauge bosons have the direction $T_{iX}={1 \over \sqrt2}
(T_{2i+7}+iT_{2i+8})$, $T_{iY}={1 \over \sqrt2}
(T_{2i+14}+iT_{2i+15})$ in gauge space. The gauge fields are
similarly related to the $A^\mu_i$. Then
\beq{
A^\mu = {1 \over \sqrt2}
\pmatrix{\ &\ &\ &\vdots&{\overline X}^\mu_1&{\overline Y}^\mu_1 \cr
         SU(3) &{\rm gauge} &{\rm fields} &
            \vdots&{\overline X}^\mu_2&{\overline Y}^\mu_2 \cr
         \ &\ &\ &\vdots&{\overline X}^\mu_3&{\overline Y}^\mu_3 \cr
         \ldots&\ldots&\ldots&\ldots&\ldots&\ldots\cr
         X^\mu_1&X^\mu_2&X^\mu_3&\vdots&SU(2)&{\rm gauge}\cr
         Y^\mu_1&Y^\mu_2&Y^\mu_3&\vdots&{\rm fields}&\ \cr}
   + Y^\mu T_Y + V^\mu T_V.}
\eeq
 From the minimal coupling (\ref{eq-b8}) it is clear
 that ${1 \over \sqrt2}
(X^\mu_i+i{\overline X}^\mu_i)$ eats the real part of $d^i_H$ and its
conjugate eats the imaginary part. Similarly, the Y-bosons eat
$u^i_H$. The V-bosons eat the $\sigma_H$-field.



\end{document}